# Institutional reforms and the employment effects of spatially targeted investment grants: The case of Germany's GRW

Björn Alecke[1] and Timo Mitze[2,*]

[1] Gesellschaft für Finanz- und Regionalanalysen (GEFRA)

[2] University of Southern Denmark, ORCID 0000-0003-3799-5200

[*] Corresponding Author: Campusvej 55, 5230 Odense, Denmark, tmitze@sam.sdu.dk

**Abstract:** Spatially targeted investment grant schemes are a common tool to support firms in lagging regions. We exploit exogenous variations in Germany's main regional policy instrument (GRW) arriving from institutional reforms to analyse local employment effects of investment grants. Findings for reduced-form and IV regressions point to a significant policy channel running from higher funding rates to increased firm-level investments and newly created jobs. When we contrast effects for regions with high but declining funding rates to those with low but rising rates, we find that GRW reforms led to diminishing employment increases. Especially small firms responded to changing funding conditions.

**Keywords:** Spatially targeted policy, lagging regions, investment grants, institutional reforms, employment, IV estimation

## 1. Introduction

This paper estimates the employment effects of Germany's key spatially targeted policy instrument – the joint task "Improvement of regional economic structure" (<u>G</u>emeinschaftsaufgabe "Verbesserung der <u>R</u>egionalen <u>W</u>irtschaftsstruktur") or simply GRW. Since its enactment in 1969 (BGBl. I, 1969), the German federal government and the *Bundesländer* (federal states) have been using the GRW to support lagging regions in managing processes of structural economic change, creating employment and strengthening overall growth. The overarching goal is to balance living conditions across all regions of Germany (Art. 72, GG). While the GRW is sometimes erroneously classified as a place-based policy, it is not a locally designed bottom-up policy but rather follows national goals which incorporate some elements of place sensitivity (Martin et al., 2022). As its first and financially most important pillar, the GRW grants investment subsidies to firms for the acquisition of machinery, the construction and modernization of buildings etc. A second pillar provides financial assistance to business-related public infrastructure projects in eglible regions. Rules for GRW support that deal with eligibility, type and intensity of funding as well as procedures for the distribution and provision of funds are defined in a coordination framework (*Koordinierungsrahmen*), which needs to be regularly approved by the federal government and states.

The coordination framework also lists requirements set by the European regional state aid rules. This is because the GRW, like all other support policies in the EU that grant investment aid to companies, is subject to the provisions of EU state aid law. This means that GRW assisted areas and maximum funding rates must be compatible with the respective applicable EU state aid law requirements, specifically the regional aid guidelines. Because guidelines on regional state aid have been periodically revised at the EU level, there have been significant changes in the GRW coordination framework in the recent past. In addition to EU state aid law requirements, the GRW coordination framework is also changed in response to gradual shifts in Germany's internal economic geography. While the GRW gave special treatment to East German regions after German re-unification over the last decades, the continued East-West income convergence throughout the 1990s and 2000s has led to the adoption of common funding rules for all German regions since 2007. As a consequence of these institutional changes to the GRW coordination framework, some regions have dropped out from funding eligibility while others have been added. Moreover, maximum funding rates for private business investments have been subject to changes over time. While, for instance, small and medium-sized enterprises (SMEs) in several West German regions saw a rise in maximum funding rates, these rates have been declining in East German urban areas over the last years.

We exploit the spatio-temporal variations in funding conditions associated with reforms to the GRW coordination framework to conduct a novel evaluation of policy effectiveness. We focus on estimating the employment effects in small and medium-sized enterprises (SMEs), which are considered as key policy outcome that shall turn the GRW instrument into a "job engine" for structurally weak regions (BMWK, 2022). Our main contribution to the available literature is to provide first evidence on GRW effectiveness in the light of its recent institutional reforms that have broadened the set of eligible regions while generally lowering maximum funding rates for firm investments. We do so by investigating overall effects for the period 2007-2020 and investigate effect heterogeneity across regions and firm-size categories.

As part of our empirical identification strategy, we exploit the quasi-experimental nature of the reforms to the GRW coordination framework since 2007 to run treatment effect regressions that additionally control for the endogeneity of investment decisions with regard to associated employment changes through instrumental variable (IV) regressions. Besides this key methodical contribution to the bulk of available GRW evaluations, we further introduce a largely unexplored data source, the GRW monitoring data, for empirical analyses of GRW effectiveness. While these data are commonly used to monitor the financial implementation of the GRW instrument, they have not been utilized for rigorous impact analyses yet. As we show, monitoring data are subject to strict quality checks and can be used as a reliable source to estimate gross employment effects of GRW funding.

In the following section, we provide a compact overview of the state-of-the-art research that has previously dealt with the firm- and region-level analysis of GRW effects on key economic outcome variables. This helps us to identify research gaps that motivate our methodical approach and embed our own empirical results within the context of earlier findings. Section 3 discusses data and conceptual issues together with a presentation of the estimation setup. In Section 4, we report the main empirical results together with estimates for the mechanisms of funding and sensitivity tests. Section 5 concludes the paper with a discussion of the expected policy effects in the aftermath of reforms to the GRW coordination framework.

## 2. Prior literature and research gaps

The economic effects of spatially targeted regional policy in Germany and, in particular, the first pillar of the GRW instrument, have been investigated in several earlier qualitative and quantitative studies. Here, we focus on a review of quantitative evaluation approaches that use econometric methods for cause-effect analyses, which have become a gold standard of evaluation in recent years at national levels and within the EU (see, e.g., Bronzini and De Blasio, 2006; Bernini and Pellegrini, 2011; de Castris and Pellegrini, 2012; Becker et al., 2012; Cerqua and Pellegrini, 2014; Decramer and Vanormelingen, 2016; Ehrlich and Overman, 2020; Aubert et al., 2022).

Earlier studies on GRW effectiveness can be grouped into 1) micro-econometric approaches, which study effects at the firm level, and 2) macro-econometric approaches, which analyze GRW effects at the aggregate (regional) level. Identification of micro effects of GRW funding at the firm level can thereby be seen as a necessary condition for the existence of regional (or some other type of aggregate) effects of policy support. Macro-econometric studies add to this by investigating the 'net' effects of funding. Larger regional 'net' effects of funding relative to the sum of micro effects identified at the firm level are typically associated with positive spillovers to non-funded firms – for example, through underlying input-output linkages, multiplier effects or positive agglomeration externalities. Smaller 'net' effects would instead indicate that GRW funding has a negative distortive effect on non-funded firm due to competition or crowding out.

First micro-econometric impact studies of the GRW instrument have been published in the early 2000s, mainly using survey data from the IAB Establishment Panel (Stierwald and Wiemers, 2003, Ragnitz, 2003, Lehmann and Stierwald, 2004). These studies, which focus on East German firms in the late 1990s and early 2000s, find that GRW funding significantly increased firm-level investment rates, while funding induced only moderate deadweight effects. Follow-up research added a focus on employment effects (Alecke et al., 2012, Bade and Alm, 2010, Bade, 2013, Brachert et al., 2020). Overall, these latter studies report a significantly positive employment development in firms in the first years after the GRW-funded investment project had been implemented.

Bade and Alm (2010) find that the employment growth of funded SMEs during 1999-2006 exceeded those of non-funded firms by approx. 11 %-points. Bade (2013) has extend this approach to an analysis of large firms with more than 250 employees. The study finds a similar relative employment effect of GRW funding of 9 %-points. Brachert et al. (2020) use difference-in-difference estimation to identify differences in employment growth between funded and non-funded firms in the first five years after the GRW-funded investment project had been implemented. Estimation results point to higher employment levels of GRW funded firms of up to 15 %-points three years after the project completion. In comparison, only small wage effects and no effects for labor productivity or changes in the export ratio were found.

Compared to micro-econometric studies, the scope, estimation setup and sample settings differ more widely in macro-econometric approaches. Nonetheless, some overall conclusions can be drawn: Prior research has reported a positive effect of GRW funding on regional investment rates (Schalk and Untiedt, 2000, for West Germany), together with increases in regional employment (Blien et al., 2003, for East Germany), per capita GDP and productivity growth (Alecke and Untiedt, 2007, Eberle et al., 2019, for West and East Germany). Mitze et al. (2015) find evidence for non-linearities in the relationship between funding intensities and productivity growth at the regional level by applying a generalized propensity score model. Studies that are concerned with 'net' effects of GRW funding in the presence of spatial spillovers report mixed results. While Eckey and Kosfeld (2005) only find weak positive ´net' effects of funding on per capita income, Alecke et al. (2013) report evidence for positive spillovers of GRW funding.

Eberle et al. (2019) have used a VAR approach to analyze the regional effects of GRW support on various regional outcomes such as per capita income development, the investment rate, employment, human capital endowment, and the regional technology level. The study also distinguishes between the effects of GRW investment support and the promotion of public infrastructure, i.e., the second pillar of the GRW. The study finds significant positive funding effects for the regional GDP per capita development, regional employment levels and human capital stock. Brachert et al. (2019) also identify positive effects in terms of gross value added and labor productivity for West German regions using a regression discontinuity approach but find no direct funding effects with regard to regional employment or wage development.

In the most recent study, Siegloch et al. (2021) use a multilevel setup combining firm- and regional level data to evaluate the employment effect of GRW funding. Similar to the approach taken here, the authors exploit changes in maximum GRW funding rates in eligible regions (only in East Germany) over the period 1996-2013 to isolate the effect of changes in maximum funding rates on outcomes using an event study design. Although, the study uses various micro data sources to measure their key outcome variables of manufacturing investment and employment at the county level (NUTS3 level), their work is essentially not a strict micro-level approach: No direct comparison of subsidized and non-subsidized units is made at the firm level because the authors are unable to identify which firm did in fact receive GRW subsidies and which did not. Instead, the available micro data are used to estimate average employment and wage effects in counties which are subject to changes in the GRW coordination framework. The authors find that a 1% reduction in GRW maximum subsidy rates at the county level leads to a 6.7% reduction in the overall investment level and 1% reduction in manufacturing sector employment after ten years. The authors do not find evidence for wage effects but add to the discussion on spillovers of GRW funding with evidence presented for intra- but not inter-regional spillovers on untreated firms/sectors.

Overall, with few exceptions, prior studies have largely focused on analyses of GRW effectiveness for a homogeneous group of funded firms/regions (mostly in East Germany) during times with low to moderate variation in underlying funding conditions. Little knowledge is avaible on threshold effects of funding and effect heterogeneity that considers variation in funding intensities or conditions (except for Mitze et al., 2015, and Siegloch et al., 2021). Identifying the impact of changing funding conditions on employment effects can provide important insights in GRW funding effectiveness. Particularly, the recent institutional reforms associated with 1) a broadened the group of regions eligible for funding together with 2) a reduction in maximum funding rates (mostly in East German regions) and 3) a simultaneous increase in funding rates in West Germany, offer an unprecedented exogenous variation in underlying GRW funding conditions. This allows us to investigate more precisely under what conditions funding translates into higher investment rates and subsequent employment. It also allows us to study effect heterogeneity across regions and firm types, i.e., micro and small firms compared to medium-sized firms.

Lastly, by focusing on the subgroup of funded regions subject to changes in funding conditions, we avoid problems associated with self-selection into treatment which typically arise in estimation settings the compare funded and non-funded firms or regions. Nonetheless, we carefully account for other sources of estimation biases, specifically, by using an instrumental variable (IV) approach to account for the endogeneity between investment and employment rates in funded firms when identifying the policy channel of investment grants running through changes in maximum funding rates as an incentive for investment decisions.

## 3. Data, concept and empirical specification

*Data and conceptual issues*

Data for the empirical analysis have been obtained from the Federal Office for Economic Affairs and Export Control (BAFA). As a higher federal authority, the BAFA is responsible for reporting and monitoring the implementation of the GRW on behalf of the Federal Ministry of Economics and Climate Protection (BMWK). One key task of the BAFA is to collect and process statistical data about funding flows and recipients, which is made available to federal authorities and for parliamentary inquiries. Different from survey data such as the IAB Betriebspanel, data collected by the BAFA has the form of a nationwide registry covering all GRW funding applications approved by federal states. For this purpose, the approval offices of the federal states involved in the GRW funding process transmit approval notices to the BAFA for statistical recording. In the course of this recording process, the BAFA checks approval notices for compliance with funding rules.

We have been provided with a detailed BAFA dataset containing all approved cases of GRW investment support to firms since 2007. The data set comprises 26,129 cases, which have been aggregated to 15,568 data records at the municipal level (*Gemeinden,* LAU2 level) by year and for two categories of size classes: 1) micro and small firms with less than 50 employees and 2) medium-sized firms up to 250 employees. The data cover information on the overall investment sum (in euro), the GRW funding amount (in euro), equity and debt capital (in euro) used by the funded firm to implement the investment project as well as the number of permanent jobs in the firm before the start of the investment project together with information on the number of permanent jobs created/retained through the investment project. Overall, this equips us with a panel data set for over 2,550 municipalities in Germany with funded GRW investment projects during the period 2007-2020. An overview of variables used in the empirical analysis together with descriptive statistics is provided in Table A1 in the appendix.

Some noteworthy aspects about the BAFA data are in order: The funding and employment information used here are recorded at a specific settlement date when the firm applies for funding. Proof of use is checked during the implementation phase of the investment project. Since firms are required to pay back funding if financial resources are not used or target values for employment are not met, applicant firms have strong incentives not to overstate the number of job created/pertained when applying for funds. A comparison of available (subsample) data between targeted and actual investment and employment numbers shows that targeted employment numbers are typically somewhat smaller than actual employment realizations. This indicates that results presented on the basis of the BAFA dataset recorded at the settlement data of funding application constitute a conservative lower bound of the overall funding effect.

As the BAFA dataset only covers municipalities eligible for funding during our sample period (non-funded German municipalities are automatically excluded from the sample), we do not aim at comparing funded and non-funded municipalities in a "with-without" setup. Rather, given the relatively long sample period between 2007 and 2020, the data offer us the possibility to compare outcomes and funding intensities among treated municipalities that exploit exogenous changes in maximum funding rates driven by reforms in the institutional setup of the GRW and associated EU state aid law for effect identification. Hence, our identification approach aims at estimating changes in the investment intensity for GRW funded projects and the associated number of jobs created through these investments associated with changes in maximum funding rates that differ at the municipal level. Overall, our sample period covers the following GRW coordination frameworks 36 (2007-08), 361 (2008-10), 362 (2010-14), 3631 (2014-17), and 3632 (2018-20).

What is important for our empirical identification strategy is that eligible regions and the regional differentiation of maximum funding rates is determined in each coordination framework at the national level by the federal government based on a composite indicator that measures the past economic performance of regions (Eckey et al., 2007; Schwengler, 2013; Maretzke et al., 2019). The latter is assessed on the basis of a region's per capita earnings, the unemployment rate and the quality of the local infrastructure relative to a benchmark level. While East and West German regions were separately assessed according to this composite indicator prior to 2007, since then all German regions are jointly assessed and ranked according to their relative backwardness to the German average. This is used to classify regions into those with high and low funding priority. As the calculation of past economic performance may change in the transition between two GRW coordination frameworks, this may also change a region's priority status and funding eligibility together with maximum funding rates. Together with the restrictions determined by changes in EU regional state aid rules as outlined above, this imposes exogenous changes to a region's funding status over time (for details see, e.g., Deutscher Bundestag, 2007, 2009, 2014).

As shown in Table 1, maximum and average GRW funding rates have generally declined during the period 2007-2020 whenever a new GRW coordination framework came into force. While maximum rates have been globally reduced from 50% to 40% (local changes could be higher), average de facto funding rates, defined as public subsidies per total fundable investment volume of an investment project, declined from 42.3% to 28.7% (for micro and small firms) and from 34.5% to 19.7% (for medium-sized firms). By contrast, the average investment volume per employee and the number of new jobs created show mixed trend developments over this time period. This is partly attributable to the fact that maximum funding rates differ considerably in range across municipalities for each GRW framework plan (Panel A in Figure 1) and have developed in different directions at the local area level with about half of the GRW funded regions experiencing a decline in funding rates of up to 20%-points, while other regions see a rise in funding rates of up to 30%-points (see Panel B in Figure 1).

Especially East German urban areas experienced a strong decline in maximum funding rates for investment projects implemented in micro and small firms. This decline is illustrated in Panel C of Figure 1 for the municipality of Zeulenroda (Thuringia) in East Germany, which experienced a decline from 50% to 30% between 2007 and 2020. At the same time, maximum funding rates in West German municipalities, such as Bochum in the Ruhr area, increased from 0% up to 30% (Panel D in figure 1) during that period. Funding rates for Zeulenroda and Bochum have thus converged during the last decade. Resulting territorial changes in the funding landscape are shown in Figure 2. Overall, the reforms of the GRW instrument have led to two key development trends: 1) the absolute level of maximum funding rates has been reduced over time, 2) reforms operated in a spatially differentiated manner with funding rates in East Germany being reduced significantly, while structurally weak West German regions have experienced rising rates.

**Table 1: Funding, investment rates and newly created jobs by GRW framework plans**

|  | FP 36 | FP 361 | FP 362 | FP 3631 | FP 3632 |
|---|---|---|---|---|---|
| **Micro and Small Firms (with <50 employees)** | | | | | |
| Maximum funding rate (in %) | 50.0 | 50.0 | 50.0 | 40.0 | 40.0 |
| ∅ funding rate* (in %) | 42.3 | 40.5 | 39.7 | 31.2 | 28.6 |
| Investment per employee (in €) | | | | | |
| ➢ Mean | 122,718.8 | 156,603.9 | 130,268.4 | 144,080 | 167,584.8 |
| ➢ Min. | 1456.25 | 0 | 0 | 0 | 200 |
| ➢ Max. | 9,564,599 | 9,497,000 | 5,195,000 | 8,859,830 | 1.26E+07 |
| Newly created jobs (total) | | | | | |
| ➢ Mean | 0.29 | 0.38 | 0.23 | 0.19 | 0.15 |
| ➢ Min. | 0 | 0 | 0 | 0 | 0 |
| ➢ Max. | 190 | 166 | 240 | 120 | 128 |
| **Medium-Sized Firms (with ≥50 to <250 employees)** | | | | | |
| Maximum funding rate (in %) | 40.0 | 40.0 | 40.0 | 30.0 | 30.0 |
| ∅ funding rate* (in %) | 34.5 | 34.3 | 31.7 | 22.6 | 19.7 |
| Investment per employee (in €) | | | | | |
| ➢ Mean | 47,648.87 | 44,199.07 | 41,496.97 | 46,348.77 | 42,595.17 |
| ➢ Min. | 852.45 | 482.35 | 0 | 0 | 29.97 |
| ➢ Max. | 390,000 | 1,188,732 | 720,114.9 | 481,914.9 | 424,461.5 |
| Newly created jobs (total) | | | | | |
| ➢ Mean | 16.89 | 11.93 | 15.48 | 10.07 | 7.67 |
| ➢ Min. | 0 | 0 | 0 | 0 | 0 |
| ➢ Max. | 139 | 241 | 282 | 100 | 194 |

*Notes:* * Computed for cases with funding rates >0; FP = GRW coordination framework plan.

*Source:* Own calculations based on data from BAFA (2021).

**Figure 1 Temporal development of maximum GRW funding rates 2007-2020**

Panel A: Evolution of maximum funding rates across municipalities and GRW framework plans

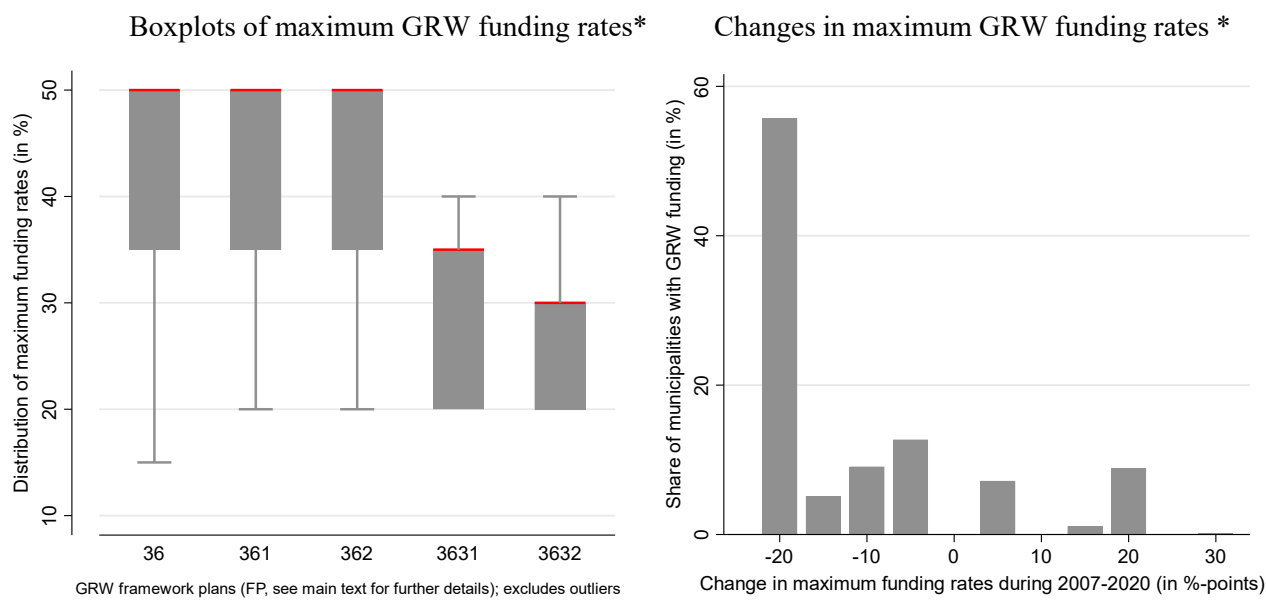

Panel B: Exemplary development of GRW funding rates in different municipalities and size classes

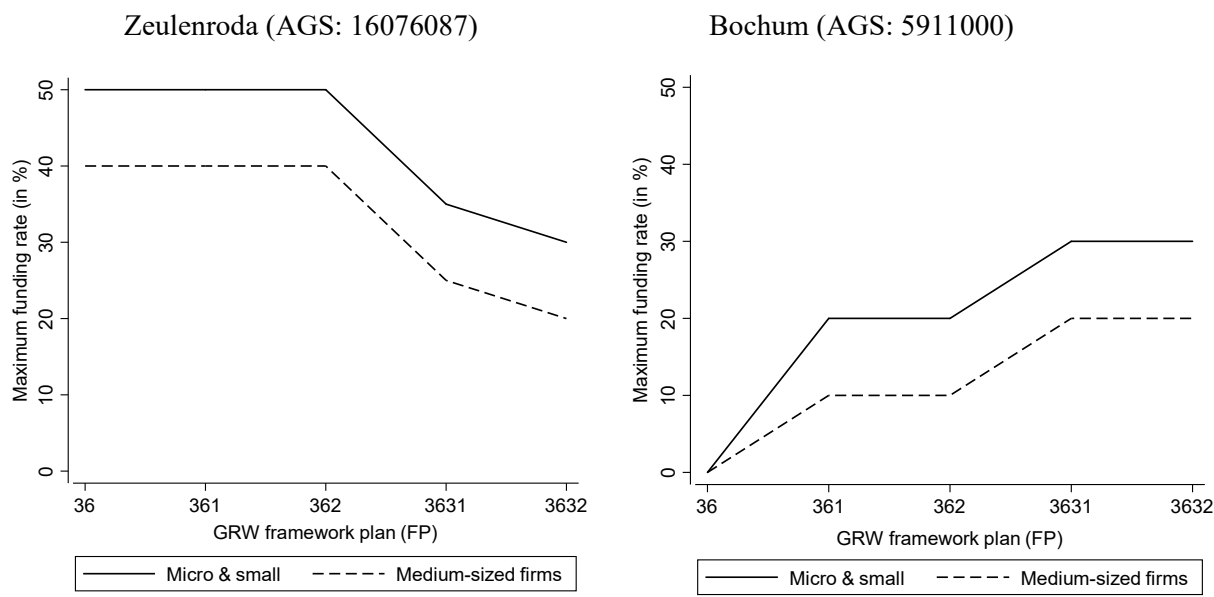

*Notes:* * Maximum GRW funding rates for micro and small firms up to 50 employees. GRW coordination framework plans 36 (years 2007-08), 361 (2008-10), 362 (2010-14), 3631 (2014-17), and 3632 (2018-20). AGS indicates the municipal code in German official statistics.

*Sources:* Own calculations based on data from BAFA (2021).

**Figure 2: Spatial distribution of maximum funding rates for GRW investment support**

Panel A: Maximum funding rates* at the start and end of the sample period 2007-2020

GRW framework plan 36 (until 2008)      GRW framework plan 3632 (since 2016)

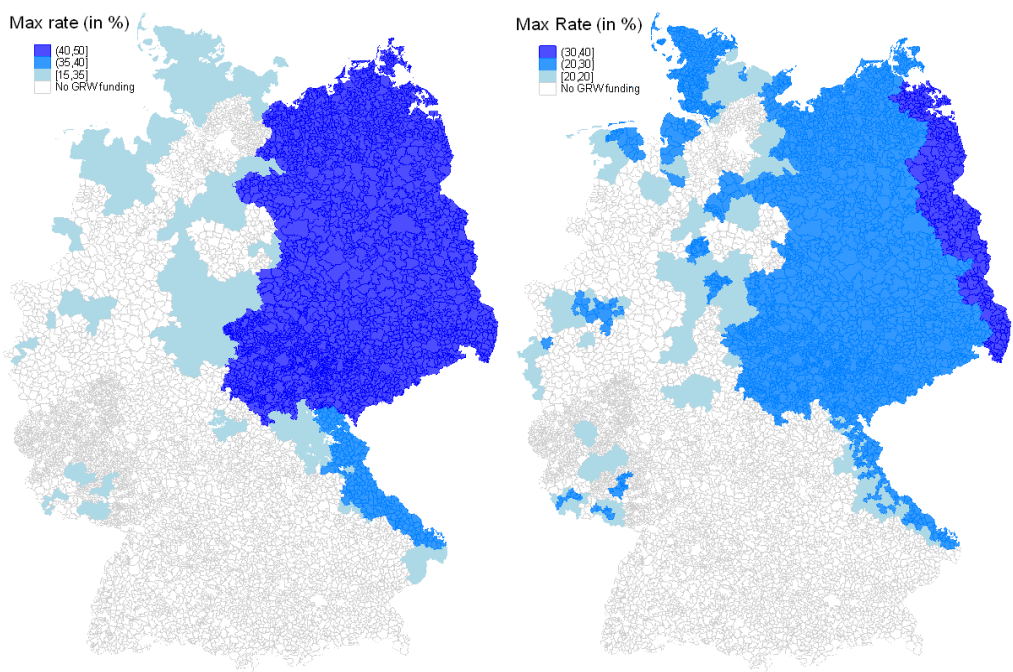

Panel B: Change in maximum fundings rates during 2007-2020 (in %-points)

Micro and Small Firms      Medium-sized Firms

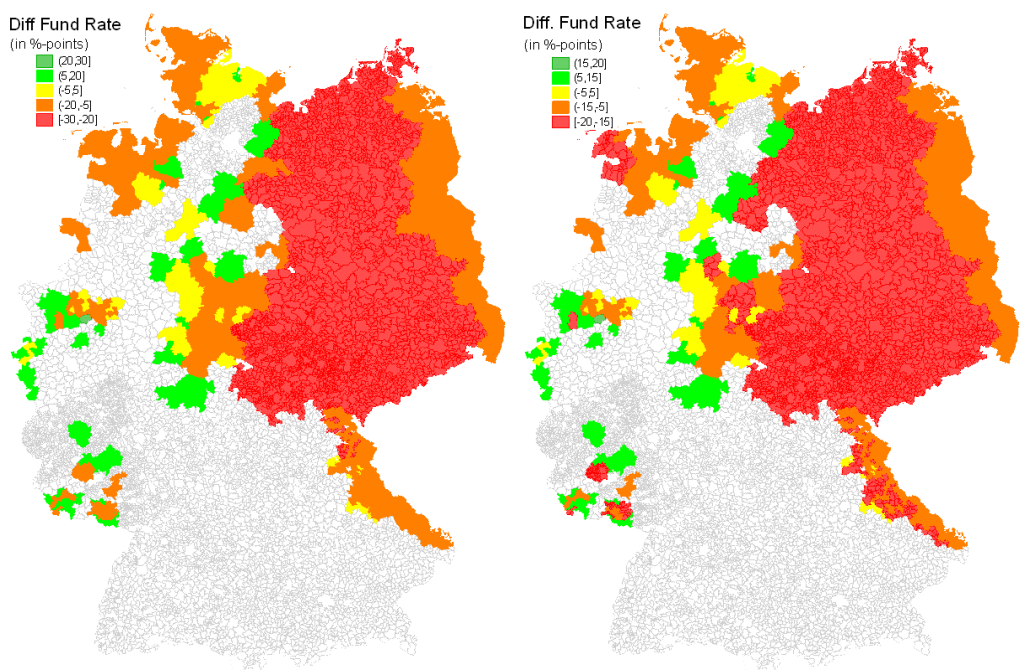

*Notes:* * Reported Maximum funding rates are those for micro and small firms up to 50 employees.
*Source:* Own calculations based on data from BAFA (2021).

*Empirical specification*

These stark changes in local funding conditions (funding rates) are exploited to investigate overall employment effects of GRW funding and explore effect heterogeneity. In order to identify the employment effects of GRW funding at the municipal level, a two-stage panel regression approach is used, which estimates the relationship between the exogenously determined maximum GRW funding rate, the investment intensity (i.e., the investment volume per employee) and the number of newly created jobs as follows:

GRW maximum funding rate $\rightarrow$ Investment intensity $\rightarrow$ Newly created jobs

Overall outcome variable is the average number of newly created jobs ($\Delta emp_{i,t+n}$) in municipality *i* at time *t+n* specified as

$$\Delta emp_{i,t+n} = \beta \cdot grw\_invpemp_{i,t} + \mu_i + \lambda_t + \sum_{m=1}^{M}(\mu_{m(i)} \times \lambda_t) + u_{i,t}, \qquad (1)$$

where $grw\_invpemp_{i,t}$ measures the average investment intensity for GRW funded investment projects in municipality *i* at time *t*, where the temporal difference between *t* and *t+n* measures the average implementation time of the investment project (which is generally set to three years according to GRW funding rules). Lowercase letters indicate that variables are log-transformed. The latter ensures that the regression parameter *β* can be interpreted as an elasticity measuring the percentage response of newly created jobs to a 1% increase in the average investment rate. Since employment effects are potentially driven by other (latent) structural factors, equation (1) controls for time-constant differences between municipalities ($\mu_i$), common time-fixed effects across municipalities to account for common business cycle trends ($\lambda_t$) and *M* macro region-specific time trends (at the federal state or county level) as $\sum_{m=1}^{M}(\mu_{m(i)} \times \lambda_t)$ ), where the interaction term $(\mu_{m(i)} \times \lambda_t)$ capture time-fixed effects for macro region *m* in which municipality *i* is located. The inclusion of this multidimensional fixed-effects (MD-FE) structure shall reduce estimation biases from latent variables such as the regional business climate that may correlated with local economic policies (e.g., Blume, 2003) so that the remainder error term $u_{i,t}$ is white noise.

While a correlation between the average investment rate and subsequent average employment changes for GRW funded firms in municipality *i* gives an important indication about policy effectiveness, some endogeneity concerns remain in order as investment and employment may be simultaneously determined as a result of firm- or municipality-specific time varying factors, which are not captured by the above-described MD-FE setup. Moreover, estimation of equation (1) does not reveal possible information about deadweight effects of funding, which would be in order if higher funding rates are not associated with an overall higher investment intensity. From an economic policy perspective, it is therefore of particular interest to find out whether spatial differences in maximum funding rates across regions and changes in the regional funding rate over time significantly affect the investment decision and subsequently employment growth at the municipal level. Hence, linking funding rates to investment and employment effects would allow us to measure if each additional euro of public funding, implied by a higher funding rate, leads to more employment via a scaling up in the investment intensity. We therefore add the following first-stage equation as

$$grw\_invpemp_{i,t} = \gamma \cdot \max\_Fundrate_{i,t} + \delta \cdot \Delta\max\_Fundrate_{i,t} + \theta' X_{i,t} + \mu_i + \lambda_t + \sum_{m=1}^{M}(\mu_m \times \lambda_t) + u_{i,t}, \qquad (2)$$

where the investment intensity ($grw\_invpemp_{i,t}$) is specified as a function of the maximum GRW funding rate ($\max\_Fundrate_{i,t}$) in municipality *i* at time *t* and the year-to-year change in the funding rate ($\Delta\max\_Fundrate_{i,t}$). As outlined above, since GRW maximum funding rates are set at the national level in a comparative assessment of the historical economic performances of larger macro regions at discrete points in time (Eckey et al., 2007; Schwengler, 2013; Maretzke et al., 2019), they can be seen as exogenous for firm-based investment decisions at the local level. Differences in maximum funding rates across regions and over time can, accordingly, be used as a source of exogenous variations in the data to identify investment and employment effects of GRW funding.

Maximum subsidy rates not only differ across space but also differ by firm size classes. We use this additional variation to estimate equations (1) and (2) separately for the group of small firms up to 49 employees and the group of medium-sized firms up to 249 employees (and more than 49 employees). If GRW support has an impact on the entrepreneurial investment decision, it is assumed that the parameter γ to be estimated is significantly greater than zero, i.e.: an increase (decrease) in the support rate over time leads to a higher (lower) investment volume per employee. Moreover, if we exclude municipal-fixed effects ($\mu_i$) in estimating equation (2), not only the difference in the maximum subsidy rate over time (within variation) but level differences in maximum funding rates between municipalities at any point in time are used for the identification of γ. Finally, it should be emphasized that we use the maximum and not the average *de facto* funding rate (i.e., the share of public funds in the total investment volume per firm) for the estimation, since the latter is composed of the investment volume and is thus partially endogenous to our outcome variable in equation (2).

The vector $\mathbf{X}_{i,t}$ includes other control variables that influence investment decisions, specifically we control for differences in external debt capital associated with GRW investments. On the one hand, a higher level of debt capital can be taken as in indication for larger investment projects which cannot be implemented using only the firms' equity and needs to undergo additional quality checks by private banks involved in this process. On the other hand, a larger share of debt capital in the total investment volume may also be associated with a higher riskiness of the project, which may be equally relevant for its expected employment effects. We estimate equation (1) as baseline specification as static panel model by OLS with MD-FE, the system of equation (1) and (2) is estimated using 2SLS. The difference between the two estimation approaches is that the 2SLS estimation only considers variations in investment rates driven by differences and changes in maximum GRW funding rates to measure the employment effects of the GRW instrument. While the 2SLS estimation is therefore more informative with regard to the effectiveness of GRW investment promotion, it should be mentioned that 2SLS is also more susceptible to estimation biases in the case of weak instruments. We therefore apply several post-estimation tests to check the robustness of the IV approach.

## 4. Empirical results

Baseline OLS regression results for micro and small firms in Table 2 show a positive and statistically significant correlation between the average investment intensity and the average number of new jobs created in municipality *i*. The estimated coefficient indicates that an increase in the investment volume per employee (already employed before the funded investment is undertaken) by 10% is accompanied by an increase in newly created jobs in funded firms by about 3%. The estimated relationship between the investment intensity and newly created jobs created is moderately higher for medium-sized enterprises (lower part of Table 2). This result is consistent with the descriptive statistics presented in Table 1, which indicate a higher employment growth for medium-sized firms on average. Overall, the estimation results presented in Table 2 are in line with earlier micro-econometric results reported in Alecke et al. (2012). However, although the OLS estimates controls for latent influence factors through the MD-FE structure (regions, time, and macro region-time FE), they do not control for the possible simultaneity between investment decisions and employment development at the firm level. This simultaneity may lead to possible biases in the reported estimation coefficients.

The 2SLS approach, accordingly, exploits variation in the investment intensity for effect identification deriving from between- and within-type changes in local funding conditions. In addition, we control for other factors, such as debt capital used, which may influence the firms' investment decision The 2SLS results in Table 2 indicate that the employment effects of funding are positive, statistically significant and larger compared to the OLS results. A 10% increase in the investment intensity is associated with an increase in newly created jobs of about 5-6%. The magnitude of employment growth thus lies between the estimated moderate effects in Alecke et al. (2012) and larger effects reported in Bade and Alm (2010) and Brachert et al. (2020), who find average employment growth of up to 10% for subsidized vis-à-vis non-subsidized firms. What needs to be considered is that the estimated positive association between maximum funding rates, a higher investment intensity and employment growth arrives from two different developments: First, it indicates that for several municipalities experiencing a decline in the maximum funding rate this translates into a lower investment intensity and fewer newly created jobs over time. Second, for regions experiencing an increase in the maximum funding rate we see a positive development in the number of jobs created running through a higher investment intensity.

**Table 2: OLS and IV regression results for employment effects of GRW funding**

| Outcome: Δemp | (1) | (2) | (3) | (4) | (5) | (6) |
|---|---|---|---|---|---|---|
| Estimation method: | OLS | OLS | OLS | 2SLS | 2SLS | 2SLS |
| **Micro and Small Firms** | | | | | | |
| grw_invpemp | 0.304*** | 0.295*** | 0.288*** | 0.625*** | 0.617*** | 0.566*** |
|  | (0.0096) | (0.0096) | (0.0121) | (0.0265) | (0.0262) | (0.0316) |
|  | | | | *First stage* | | |
| max_Fundrate | | | | 0.245 | 0.331 | 0.680 |
|  | | | | (0.3275) | (0.4959) | (0.7366) |
| Δmax_Fundrate | | | | 0.422 | 0.453 | 1.067** |
|  | | | | (0.4291) | (0.4339) | (0.4826) |
| debt_capital | | | | 0.578*** | 0.570*** | 0.520*** |
|  | | | | (0.0303) | (0.0309) | (0.0365) |
| share_debtcap | | | | −0.040*** | −0.038*** | −0.035*** |
|  | | | | (0.0033) | (0.0033) | (0.0038) |
| Obs. | 5968 | 5968 | 5968 | 5968 | 5968 | 5968 |
| Region FE | State | District | Municipality | State | District | Municipality |
| Time FE | Yes | Yes | Yes | Yes | Yes | Yes |
| (Macro-region × time) FE | Yes (State) | (State) | Yes (State) | Yes (State) | (State) | Yes (State) |
| Cragg-Donald Wald $F$ | | | | 204.44 | 199.55 | 134.48 |
| Sargan statistic | | | | 3.71 | 3.97 | 4.06 |
| ($p$ value) | | | | (0.29) | (0.27) | (0.26) |
| **Medium-Sized Firms** | | | | | | |
| grw_invpemp | 0.409*** | 0.413*** | 0.420*** | 0.511*** | 0.539*** | 0.510*** |
|  | (0.0295) | (0.0303) | (0.0384) | (0.0514) | (0.0500) | (0.0703) |
|  | | | | *First stage* | | |
| max_Fundrate | | | | −0.055 | 0.371 | 1.308 |
|  | | | | (0.5509) | (0.9696) | (1.5508) |
| Δmax_Fundrate | | | | 0.591 | 0.735 | 0.246 |
|  | | | | (0.6869) | (0.7286) | (0.9457) |
| debt_capital | | | | 0.336*** | 0.331*** | 0.274*** |
|  | | | | (0.0455) | (0.0404) | (0.0521) |
| share_debtcap | | | | −0.032*** | −0.033*** | −0.024*** |
|  | | | | (0.0065) | (0.0055) | (0.0071) |
| Obs. | 1739 | 1739 | 1739 | 1739 | 1739 | 1739 |
| Region FE | State | County | Municipality | State | County | Municipality |
| Time FE | Yes | Yes | Yes | Yes | Yes | Yes |
| (Macro-region × time) FE | Yes (State) | (State) | Yes (State) | Yes (State) | (State) | Yes (State) |
| Cragg-Donald Wald $F$ | | | | 135.84 | 126.09 | 67.81 |
| Sargan statistic | | | | 18.50 | 16.59 | 2.99 |
| ($p$ value) | | | | (0.00) | (0.00) | (0.39) |

*Notes:* ***, **, * denote statistical significance at the 1, 5, 10% critical level; robust standard errors clustered at the municipal level are given in brackets; State = German Federal State (NUTS1), County = NUTS3 region (*Kreis*), FE = Fixed effects. Further notes on the reported post-estimation tests for instrument relevance and exogeneity are given in the main text.

With regard to the effect difference between OLS and IV results two comments are in order: First, this difference points to the inconsistency of the OLS estimates, which is also confirmed by a Durbin-Wu-Hausman test. Second, it needs to be considered that the IV estimates presented here result from a discrete change in maximum funding rates for the group of funded firms. Thus, similar to Siegloch et al. (2021), our results for the estimation of the 2SLS approach in Table 2 show that employment effects can be attributed to changes in the maximum funding rate, which run through exogenously determined changes in the investment intensity. However, compared to Siegloch et al (2021), who find that a very high treatment effect of a 1% decrease in manufacturing employment (in funded and non-funded firms in a county) stemming from a 1 %-point reduction in GRW funding rates, our 0.6% estimate for the employment effect in firms with GRW-funded investment projects can be regarded as a more moderate treatment effect given that only a fraction of regional firms receive GRW funding.

The first-stage regression results in Table 2 show that we only find a significant link between changes in the maximum funding rate and the investment intensity if we control for the full set of municipality-fixed effects (i.e., in column (6)). The results also show that controlling for the level and share of debt capital in the overall investment volume is important. Post-estimation tests for the goodness of fit of the 2SLS estimation presented in Table 2 show that the instruments are 1) relevant (as measured by the reported Cragg-Donald Wald F statistic compared to the critical values of Stock and Yogo, 2005) and 2) exogenous with respect to the second stage outcome variable (as measured by the statistically insignificant values of the Sargan test). The 2SLS estimation results for medium-sized enterprises reveal a similarly high correlation between investment level and employment growth, even though the correlation between maximum funding rates, changes in the maximum subsidy rate and investment volume per employee is weaker compared to micro and small firms.

As outlined above, the positive association between funding, investment and employment is determined by two opposing developments over time. To further investigate the issue of effect heterogeneity between municipalities experiencing rising and falling maximum fund rates, we split the sample into groups along two different criteria. First, in columns (1) and (2) of Table 3 municipalities are grouped into 1) those experiencing an increase in maximum funding rates (several West German regions, see Figure 2) and 2) those with overall decreasing maximum funding rate over time (mainly East German regions, but also border regions in Bavaria, among others). Furthermore, in columns (3) and (4) of Table 3, municipalities are divided along the maximum absolute GRW funding rate into two groups with rates below and above the 75th percentile for the overall distribution of maximum GRW funding rates (computed separately for micro and small firms as well as medium-sized firms). The cut-off level was chosen to produce two roughly equal-sized subsamples, while accounting for the left-skewed distribution of maximum fund rates. As the estimation results in Table 3 show, both subsampling approaches produce very similar results.

For micro and small enterprises as well as for medium-sized enterprises, it is noticeable that the link between maximum funding rates, investment intensities and newly created jobs is significant and sizable for municipalities that belong to the group with the highest absolute funding rates while experiencing a decline in rates over time (see column (2) and column (4) in Table 3). These are mainly East German municipalities and West German municipalities located in state-level border regions in Bavaria and Lower Saxony. These municipalities initially benefitted from initial maximum funding rates of up to 50% of the overall investment sum, which have, however, declined by up to 20 %-points between 2007 and 2020. In comparison, the estimated employment effect is weaker or absent for the subsample of municipalities experiencing rising but generally lower funding rates. Especially for medium-sized firms no significant link is found here (see Table 3).

To test whether this effect is determined by differences in the underlying economic structure in East and West Germany or, in fact, by changing GRW funding conditions, Table A2 in the appendix reports additional subsample estimates, which split the sample into East and West German municipalities without distinguishing between absolute funding rates and their development over time. The result show that the overall size of treatment effects is very similar for both subgroups, which can be taken as an indication that differences in employment effects as identified in Table 3 are rather determined by changes in funding regimes rather than confounding, underlying macro-regional economic differences between East and West Germany.

**Table 3: Subsample estimates on effect heterogeneity of changes in GRW funding conditions**

| Outcome: Δemp | (1) | (2) | (3) | (4) |
|---|---|---|---|---|
| Sample: | Δmax_Fundrate ≥0 | Δmax_Fundrate <0 | max_Fundrate <75pct. | max_Fundrate ≥75pct. |
| Estimation method: | 2SLS | 2SLS | 2SLS | 2SLS |
| **Micro and Small Firms** | | | | |
| grw_invpemp | 0.321** | 0.638*** | 0.269 | 0.676*** |
|  | (0.1269) | (0.0408) | (0.2472) | (0.0445) |
| *First stage* | | | | |
| max_Fundrate | -0.395 | 1.959** | -1.011 | 2.790** |
|  | (0.2589) | (0.8928) | (3.0263) | (1.2102) |
| Δmax_Fundrate | 1.042 | 1.045* | 0.350 | 1.168* |
|  | (1.7052) | (0.5982) | (0.9603) | (0.6751) |
| debt_capital | 0.166 | 0.537*** | 0.273** | 0.547*** |
|  | (0.1412) | (0.0301) | (0.1060) | (0.0329) |
| share_debtcap | -0.008 | -0.036*** | -0.014* | -0.039*** |
|  | (0.0099) | (0.0028) | (0.0083) | (0.0031) |
| Obs. | 776 | 5192 | 2022 | 3946 |
| Region FE | Municipality | Municipality | Municipality | Municipality |
| Time FE | Yes | Yes | Yes | Yes |
| (Macro-region × time) FE | Yes (State) | Yes (State) | Yes (State) | Yes (State) |
| Cragg-Donald Wald $F$ | 1.482 | 83.174 | 1.999 | 71.462 |
| Sargan statistic | 10.64 | 11.57 | 2.20 | 8.36 |
| ($p$ value) | (0.91) | (0.01) | (0.53) | (0.04) |
| **Medium-Sized Firms** | | | | |
| grw_invpemp | 0.068 | 0.517*** | 0.271 | 0.517*** |
|  | (0.4433) | (0.1278) | (0.4319) | (0.1294) |
| *First stage* | | | | |
| max_Fundrate | 1.850 | 1.140 | -0.303 | 1.777 |
|  | (3.1853) | (1.7367) | (3.7196) | (1.8370) |
| Δmax_Fundrate | -2.099 | 0.289 | 0.781 | 0.243 |
|  | (3.3568) | (0.9513) | (2.4601) | (0.9483) |
| debt_capital | 0.086 | 0.275*** | 0.335 | 0.270*** |
|  | (0.2739) | (0.0340) | (0.2034) | (0.0340) |
| share_debtcap | -0.017 | -0.024*** | -0.019 | -0.023*** |
|  | (0.0344) | (0.0039) | (0.0218) | (0.0039) |
| Obs. | 317 | 1576 | 524 | 1540 |
| Region FE | Municipality | Municipality | Municipality | Municipality |
| Time FE | Yes | Yes | Yes | Yes |
| (Macro-region × time) FE | Yes (State) | Yes (State) | Yes (State) | Yes (State) |
| Cragg-Donald Wald $F$ | 0.46 | 16.61 | 0.98 | 16.15 |
| Sargan statistic | 22.27 | 6.70 | 1.95 | 5.88 |
| ($p$ value) | (0.02) | (0.08) | (0.58) | (0.12) |

*Notes:* ***, **, * denote statistical significance at the 1, 5, 10% critical level; robust standard errors clustered at the municipal level are given in brackets; State = German Federal State (NUTS1), FE = Fixed effects. Further notes on the reported post-estimation tests for instrument relevance and exogeneity are given in the main text.

This empirical picture is furthermore underlined when we run leave-one-out sensitivity checks. These checks sequentially leave out one federal state from the sample in order to see if the estimated treatment effects remain stable or break down in their statistical significance and size. In the latter case, this would be a strong indication that effects are only driven by the local development in a specific state rather presenting a general association between GRW funding conditions, investment and employment changes across all German municipalities that have received the treatment. As the results in Table A4 show, though, the effects are quite robust to such sample reductions. If at all, we find somewhat larger employment effects once we exclude municipalities in the federal state of Saxony (Sachsen) from the sample, particularly for the group of micro and small firms. As shown in Table A5, Saxony is the state that was mostly affected by institutional changes in the GRW in terms of the number of municipalities that have experienced decreasing maximum funding rates between 2007 and 2020. Hence, the leave-one-out robustness tests lend further support to the hypothesis that heterogeneous employment effects largely result from the different regional trajectories mirroring decreasing and increasing funding rates over time.

This observed effect heterogeneity can, hence, be interpreted in such a way that the reform-based reductions in maximum GRW funding rates (mainly in urban areas in East Germany) have led to declining investment intensities over time and that these lower investment rates have, in turn, translated in lower employment effects. Siegloch et al. (2021) come to a similar conclusion. This points to two implications: First, the development of maximum funding rates over time is relevant for investment demand and the creation of new jobs by firms. Second, the absolute level of the subsidy rate also plays a role in determining whether companies invest in the construction of new plants or the expansion of existing capacities. Thus, the estimation results for West German regions (columns (1) and (3) in Table 3), which have experienced increasing maximum funding rates but only to moderate absolute subsidy rates (max. 30%), show a weaker or even statistically insignificant correlation between maximum GRW funding rates, average investment volume per employee and newly created jobs in a municipality. Especially micro and small firms have been found responsive to changes in GRW funding conditions, which point to the role of capital subsidies for these firms.

## 5. Discussion and conclusion

The GRW is the main policy instrument to support the economic development in lagging regions in Germany through investment grants to firms. While GRW investment support had largely focused to support East-West income convergence throughout the 1990s and 2000s, several reforms have altered funding conditions for small and medium-sized firms in West and East Germany. While the number of the number of regions eligible for funding has been increased to include structurally weak regions in West Germany, maximum funding rates for investment projects have generally declined over time. These two trends implied that East German regions, particularly urban areas, saw a significant reduction in maximum GRW funding rates, while firms in lagging regions in West Germany experienced rising funding rates for investments. We have used the spatio-temporal variations in funding conditions arriving from exogenous decisions at the national and EU-level to provide novel estimates about the overall employment effects of GRW investment support to eligible German municipalities (LAU2 level) during the time period 2007-2020. We have also addressed the issue of effect heterogeneity across regions and firm types to gain insights on the mechanisms of local economic effects arriving from investment grants.

Regression results for detailed GRW monitoring data show that changes in maximum funding rates translate into changes in the average investment intensity of firms and ultimately result in newly created jobs associated with these investments. We find that micro and small firms with less than 50 employees respond more strongly to changes in underlying funding conditions and they particularly do so in municipalities that saw a decline in funding rates over time. In comparison, employment effects in municipalities expiring low but rising funding rates are smaller or absent. This points to potential threshold effects of funding. Through a series of robustness tests, we rule out that the identified effect asymmetry between municipalities with high but declining vis-à-vis those with rising but moderate absolute funding rates is simply a reflex of different underlying economic conditions in East and West Germany. Rather, variations in maximum funding rates in the light of institutional reforms of the GRW coordination framework can be attributed to this effect asymmetry. Taken together, our results indicate that the recent spatio-institutional reforms of the GRW instrument, which have increased the number of regions eligible for funding receipt while lowering absolute funding rates, likely translate into smaller employment effects in the future.

# Appendix

**Table A1: Variable definitions and descriptive statistics**

| Micro and small firms up to 49 employees | | Obs. | Mean | S.D. | Min. | Max. |
|---|---|---|---|---|---|---|
| $\Delta emp_{i,t}$ | Newly created employment in GRW funded investment projects (in total persons) | 5,968 | 4.49 | 8.52 | 0 | 240 |
| $emp_{i,t}$ | Number of employees in firms receiving GRW funding (in total persons) | 5,968 | 17.37 | 13.05 | 1 | 49 |
| $grw\_inv_{i,t}$ | Investment volume of GRW funded projects (in million Euro) | 5,968 | 1.43 | 2.90 | 0 | 67.50 |
| $grw\_invpemp_{i,t}$ | GRW investment volume per employee (in million Euro per employee) | 5,968 | 0.14 | 0.47 | 0 | 12.58 |
| $max\_Fundrate_{i,t}$ | Maximum GRW funding rate (in %) | 5,968 | 37.61 | 10.92 | 0 | 50 |
| $\Delta max\_Fundrate_{i,t}$ | Annual change in maximum GRW funding rate (in %-points) | 5,968 | -0.66 | 3.55 | -16 | 20 |
| $debt\_capital_{i,t}$ | Debt capital per GRW funded investment project (in million Euro) | 5,968 | 0.05 | 0.24 | 0 | 10.24 |
| $share\_debtcap_{i,t}$ | Share of debt capital in total investment volume of GRW funded projects (in %) | 5,968 | 5.77 | 11.90 | 0 | 129.75 |
| **Medium-sized firms with 50 to 249 employees** | | | | | | |
| $\Delta emp_{i,t}$ | Newly created employment in GRW funded investment projects (in total persons) | 1,739 | 12.20 | 18.94 | 0 | 282 |
| $emp_{i,t}$ | Number of employees in firms receiving GRW funding (in total persons) | 1,739 | 106.08 | 46.50 | 50 | 249 |
| $grw\_inv_{i,t}$ | Investment volume of GRW funded projects (in million Euro) | 1,739 | 4.53 | 5.96 | 0 | 84.4 |
| $grw\_invpemp_{i,t}$ | GRW investment volume per employee (in million Euro per employee) | 1,739 | 0.04 | 0.06 | 0 | 1.19 |
| $max\_Fundrate_{i,t}$ | Maximum GRW funding rate (in %) | 1,739 | 38.31 | 10.43 | 0 | 50 |
| $\Delta max\_Fundrate_{i,t}$ | Annual change in maximum GRW funding rate (in %-points) | 1,739 | -0.69 | 3.46 | -15 | 20 |
| $debt\_capital_{i,t}$ | Debt capital per GRW funded investment project (in million Euro) | 1,739 | 0.22 | 0.73 | 0 | 10.90 |
| $share\_debtcap_{i,t}$ | Share of debt capital in total investment volume of GRW funded projects (in %) | 1,739 | 6.72 | 13.77 | 0 | 165.14 |

*Notes:* Variables are constructed on the basis of data obtained from the BAFA (2021).

**Table A2: Subsample estimates for East and West German municipalities**

| Outcome: $\Delta emp$ | (1) | (2) | (3) | (4) |
|---|---|---|---|---|
| Sample: | *East Germany* | *West Germany* | *East Germany* | *West Germany* |
| Estimation method: | 2SLS | 2SLS | 2SLS | 2SLS |
| | **Micro and Small Firms** | | **Medium-sized Firms** | |
| grw_invpemp | 0.622*** | 0.713*** | 0.573*** | 0.637*** |
| | (0.0414) | (0.1354) | (0.1548) | (0.2205) |
| Obs. | 3276 | 2692 | 1061 | 678 |
| Region FE | Municipality | Municipality | Municipality | Municipality |
| Time FE | Yes | Yes | Yes | Yes |
| (Macro-region × time) FE | Yes (State) | Yes (State) | Yes (State) | Yes (State) |

*Notes:* ***, **, * denote statistical significance at the 1, 5, 10% critical level; standard errors clustered at the municipal level are given in brackets; State = German Federal State (NUTS1), FE = Fixed effects.

**Table A3: Share of municipalities with declining maximum fund rates 2007-2020**

| Federal state | East Germany | Municipalities with increasing max. funding rate 2020 to 2007 | Municipalities with decreasing max. funding rate 2020 to 2007 |
|---|---|---|---|
| Schleswig-Holstein | | 64 | 106 |
| Niedersachsen | | 76 | 199 |
| Bremen | | 0 | 2 |
| Nordrhein-Westfalen | | 79 | 15 |
| Hessen | | 39 | 42 |
| Rheinland-Pfalz | | 38 | 48 |
| Bayern | | 7 | 217 |
| Saarland | | 15 | 10 |
| Berlin | Y | 0 | 1 |
| Brandenburg | Y | 0 | 270 |
| Mecklenburg-Vorpommern | Y | 0 | 316 |
| Sachsen | Y | 0 | 351 |
| Sachsen-Anhalt | Y | 0 | 153 |
| Thüringen | Y | 0 | 220 |

*Notes:* Variables are constructed on the basis of data obtained from the BAFA (2021). Calculations are based on changes in maximum funding rates for micro and small firms.

29**Table A4: Leave-one out subsample estimates for German federal states**

| Outcome: $\Delta emp$ | (1) | (2) |
|---|---|---|
| Sample: | *Micro and Small Firms* | *Medium-sized Firms* |
| Estimation method: | 2SLS | 2SLS |
| excl. Schleswig-Holstein | 0.566*** | 0.502*** |
| | (0.0319) | (0.0758) |
| excl. Niedersachsen | 0.566*** | 0.511*** |
| | (0.0321) | (0.0748) |
| excl. Bremen | 0.566*** | 0.511*** |
| | (0.0322) | (0.0749) |
| excl. Nordrhein-Westfalen | 0.570*** | 0.506*** |
| | (0.0322) | (0.0749) |
| excl. Hessen | 0.566*** | 0.510*** |
| | (0.0322) | (0.0749) |
| excl. Rheinland-Pfalz | 0.562*** | 0.515*** |
| | (0.0321) | (0.0750) |
| excl. Bayern | 0.569*** | 0.522*** |
| | (0.0321) | (0.0757) |
| excl. Saarland | 0.562*** | 0.506*** |
| | (0.0320) | (0.0748) |
| excl. Berlin | 0.566*** | 0.510*** |
| | (0.0322) | (0.0750) |
| excl. Brandenburg | 0.544*** | 0.535*** |
| | (0.0361) | (0.0767) |
| excl. Mecklenburg-Vorpommern | 0.561*** | 0.493*** |
| | (0.0340) | (0.0817) |
| excl. Sachsen | 0.628*** | 0.444*** |
| | (0.0433) | (0.0923) |
| excl. Sachsen-Anhalt | 0.543*** | 0.506*** |
| | (0.0348) | (0.0757) |
| excl. Thüringen | 0.568*** | 0.557*** |
| | (0.0345) | (0.0926) |
| Region FE | Municipality | Municipality |
| *Time FE* | Yes | Yes |
| *(Macro-region × time) FE* | Yes (State) | Yes (State) |

*Notes:* ***, **, * denote statistical significance at the 1, 5, 10% critical level; standard errors clustered at the municipal level are given in brackets; State = German Federal State (NUTS1), FE = Fixed effects.